\documentclass[journal=ancac3,manuscript=article]{achemso}

\usepackage{xcolor}
\usepackage{graphicx}
\usepackage{siunitx}

\definecolor{aureolin}{rgb}{0.99, 0.93, 0.0}

\usepackage{xr}
\makeatletter
\newcommand*{\addFileDependency}[1]{
  \typeout{(#1)}
  \@addtofilelist{#1}
  \IfFileExists{#1}{}{\typeout{No file #1.}}
}
\makeatother
\newcommand*{\myexternaldocument}[1]{%
    \externaldocument{#1}%
    \addFileDependency{#1.tex}%
    \addFileDependency{#1.aux}%
}
\myexternaldocument{SI}

\title{Automated Structure Discovery for Scanning Tunneling Microscopy}
\author{Lauri Kurki}
\affiliation[Aalto University]
{Department of Applied Physics, Aalto University, 00076 Aalto, Espoo, Finland}
\author{Niko Oinonen}
\affiliation[Aalto University]
{Department of Applied Physics, Aalto University, 00076 Aalto, Espoo, Finland}
\alsoaffiliation{Nanolayers Research Computing Ltd., London N12 0HL, United Kingdom}
\author{Adam S. Foster}
\email{adam.foster@aalto.fi}
\affiliation[Aalto University]
{Department of Applied Physics, Aalto University, 00076 Aalto, Espoo, Finland}
\alsoaffiliation[Kanazawa University]
{WPI Nano Life Science Institute (WPI-NanoLSI), Kanazawa University, Kakuma-machi, Kanazawa 920-1192, Japan}

\begin{document}
    
	
	\begin{abstract}
	\noindent
        Scanning tunnelling microscopy (STM) with a functionalized tip apex reveals the geometric and electronic structure of a sample within the same experiment. However, the complex nature of the signal makes images difficult to interpret and has so far limited most research to planar samples with a known chemical composition. Here, we present automated structure discovery for STM (ASD-STM), a machine learning tool for predicting the atomic structure directly from an STM image, by building upon successful methods for structure discovery in non-contact atomic force microscopy (nc-AFM). We apply the method on various organic molecules and achieve good accuracy on structure predictions and chemical identification on a qualitative level, while highlighting future development requirements to ASD-STM. This method is directly applicable to experimental STM images of organic molecules, making structure discovery available for a wider SPM audience outside of nc-AFM. This work also opens doors for more advanced machine learning methods to be developed for STM structure discovery.
    \end{abstract}
	

    \section*{Introduction}
    
    Scanning probe microscopy (SPM) methods are powerful tools for studying nanoscale systems with atomic resolution. As the basis of SPM methods, scanning tunnelling microscopy (STM) \cite{Binnig1982} and atomic force microscopy (AFM) \cite{Binnig1986} have been widely utilized in the characterization of various systems, such as biological samples, hybrid inorganic-organic interfaces and individual steps of on-surface reactions \cite{Forsburg1990, Pawlak2019, Dufrene2017ImagingBiology, Muller2008, Gobbi2018, Sun2022, Gross2018}. To enhance the spatial accuracy in the characterization, the probe tip can be functionalized using a chemically inert, flexible apex (often a CO molecule) to allow scanning at a very close tip-sample distance where Pauli repulsion is the dominating interaction \cite{Gross2009, Bartels1997}. With respect to the detailed characterization of atomic structures, there exists many demonstrations of the improved spatial resolution of AFM scanning with functionalized tips \cite{Gross2018, Zhong2020, Jelinek2017}, but STM is now also being increasingly used in this bond-resolved mode \cite{Jelinek2017, Sun2022, Pavlicek2017, Seibel2023, Vilas-Varela2023, Martinez-Castro2022, Song2020}. STM in particular benefits significantly from tip functionalization as sharp sub-molecular features appear in the image, revealing both the molecular skeleton and the electronic structure in high detail within the same experiment -- this is impossible using a bare metal tip \cite{Weiss2010} or by high-resolution AFM. In addition to atomic structural characterization, STM with a functionalized tip offers interesting approaches for electronic structure characterization via frontier orbitals of the tip apex \cite{Gross2011ptip} by improving the visibility of molecular orbitals \cite{Neel2023} and by distinguishing nearby molecular states \cite{Martinez-Castro2022}.

    In terms of structure characterization, the main limitations of STM are its incapability to see beyond the closest atom to the tip and the insurmountable problem of chemical identification of atoms. In fact, research using bond-resolved STM has so far been mostly limited to planar molecules consisting of only \textit{a priori} known atomic species, and the recognition of an unknown sample can require an extensive search through all possible molecules and configurations. The same problems have been experienced in the field of non-contact AFM where recent advances in machine learning image analysis have been proven effective in structure discovery and chemical identification of single molecules and ice structures \cite{Alldritt2020, Carracedo-Cosme2021, Oinonen2022a, Oinonen2022b, Tang2022, Carracedo-Cosme2023}. For STM, advanced image analysis methods have so far not been developed for bond-resolved imaging but instead the focus has been on, \textit{e.g.}, defect detection \cite{Wang2020}, molecule keypoint detection \cite{Yuan2023}, surface characterization \cite{Choudhary2021} and atom manipulation \cite{Chen2022} as well as autonomous experiments \cite{Rashidi2018}. Machine learning methods have also been used for automating the tip conditioning \cite{Wang2021} and tip functionalization \cite{Alldritt2022} in STM. It is clear that the focus in general has been on larger scale systems and processes, and there is a need for automated characterization methods for bond-resolved imaging with a CO tip.
    
    In this work, we present automated structure discovery for STM (ASD-STM), a machine learning approach for structure characterization directly from experimental bond-resolved STM images. With ASD-STM, we offer a solution to the problems in STM structure characterization while bridging the gap between image analysis methods for high-resolution AFM and STM. As STM with a functionalized tip is a more easily accessible characterization method compared to non-contact AFM \cite{Gross2011review}, this work also makes automated structure discovery available for a wider SPM audience and opens a door to more sophisticated methods to be developed for sample recognition in STM.

        
    \section*{Methods}



    \subsection*{Simulated STM images}

    \begin{figure}
        \centering
        \includegraphics[width=1\linewidth]{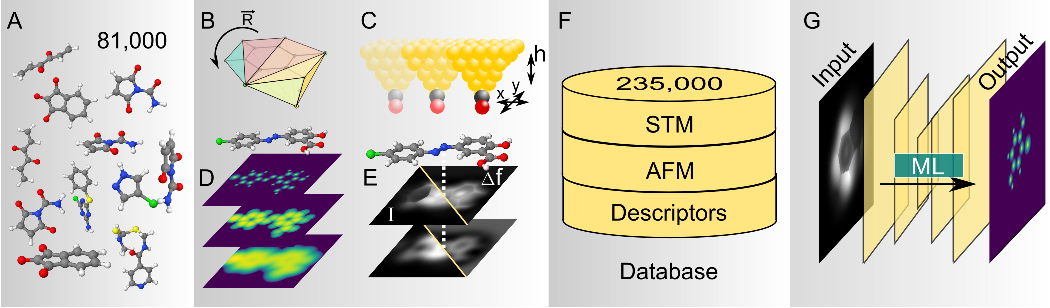}
        \caption{Scheme of ASD-STM. (A) Starting from a large set of organic molecules, (B-E) we simulate STM and AFM for different rotations of the molecules and calculate corresponding physical descriptors. (F) The images are stored in a database and (G) used as training material for a machine learning model for STM structure discovery. }
        \label{fig:scheme}
    \end{figure}
    
    The proposed method starting from a data set of molecule geometries and ending in a machine learning model that can predict atomic structure directly from experimental images is outlined in Fig. \ref{fig:scheme}. The initial data set contains the optimized geometries and atomic point charges of approximately 81,000 small organic molecules with chemical species lighter than bromine ($Z \leq 35$) with most emphasis on hydrogen, carbon, nitrogen and oxygen atoms (Fig. \ref{fig:scheme}A) \cite{Oinonen2022b}. Simulating STM images requires that we first obtain the densities of states of all molecules in the data set, and since the molecular structures are already optimized, a single-point calculation is performed for all molecules. All ab-initio calculations are performed using the FHI-aims code \cite{Blum2009, Havu2009} with the PBE functional \cite{Perdew1996} considering only the $\Gamma$ $k$-point. The unit cell is constructed by padding the molecule with $7$ Å of empty space on each side to ensure that the periodic images do not interact and affect the electronic structure. 
    
    The STM simulations for all molecules are done using the PPSTM code \cite{Krejci2017-ppstm}. PPSTM is based on the Bardeen tunnelling theory \cite{Bardeen1961} and it simulates constant height raster scanning by calculating the tunnelling current at each point of a three-dimensional grid of tip positions allowing for a quick simulation of multiple constant height scans at different tip-sample distances (Fig. \ref{fig:scheme}C,E). In PPSTM, the tunnelling current is composed of contributions from individual states of the tip and the sample. The sample states are obtained from DFT calculations as mentioned, and they are represented by  Lorentzian distributions with $0.1$ eV broadening. The tip orbitals are considered as $s$ and $p$ orbitals located at the tip apex and they provide independent tunnelling channels. The tunnelling channels are given a weight between $0$ and $1$ and the correct weights for a certain tip apex are found by matching with experiments (\textit{e.g.} approx. $10$ \% $s$ and $90$ \% $p$ orbitals for CO tip \cite{DeLaTorre2017, Sun2022}). Parallel to the STM simulations, we run the PPAFM \cite{Hapala2014-ppafm1, Hapala2014-ppafm2} code to calculate relaxation of the CO molecule near the sample and the relaxed positions are, in turn, used in the STM simulation to achieve good resemblance to experimental images revealing the molecular skeleton. PPAFM calculates the CO relaxation by considering the CO molecule as a probe particle (PP) and using a mechanical model taking into account the Lennard-Jones and electrostatic potentials between the PP and the sample. The position of the PP approaching the sample is relaxed by minimizing the total force acting upon the PP, and while calculating the relaxed tip positions, we simultaneously obtain AFM images of the molecules with little additional cost by integrating the forces affecting the PP over its path \cite{Giessibl2001}. The AFM images are included in the data set of synthetic images and they could be used as additional signal in a method developed for simultaneous AFM/STM imaging.
    
    To maximize the amount of information in the SPM images, we emphasized flat regions normal to the scanning direction and also assured a more even distribution of elements in the resulting images. To this end, we subjected the molecules to a set of rotations calculated separately for each molecule (Fig.~\ref{fig:scheme}B). The rotations were obtained by computing the convex hull of the atomic coordinates which defines planar segments of the molecule, and the rotations were weighted to include an even distribution of elements within $0.7$ Å of the plane. After applying the rotations, our data set consists of synthetic SPM images of approx. 48,000 unique molecules and 235,000 unique images (Fig. \ref{fig:scheme}F). This procedure and the full composition of the data set are described in full detail in a previous work by Oinonen \textit{et al.} \cite{Oinonen2022b}.

    One example item from the generated data set is shown in Fig. \ref{fig:scheme}D-E. All items in the data set include a stack of 10 constant height STM and AFM images taken at $0.1$ Å separation, over a range of tip-sample distances starting from a close distance scan where the molecular skeleton is revealed to a distance where only the electronic structure at the Fermi level is seen. A stack of images is created as opposed to just one slice because the real tip-sample distance in the experiment is not known, and we want to avoid overfitting to an exact scanning height by letting the model see images taken at a range of different scanning heights. This is achieved by randomly selecting one 2D slice from the 3D stack of images during training. Consequently, our model does not require a full stack of images at prediction time, making structure discovery easily accessible by using only one image. The data set also includes three image descriptors for each molecule -- atomic disks, van der Waals spheres and height map \cite{Alldritt2020} (Fig. \ref{fig:scheme}D) -- which we have chosen as visual and two-dimensional representations of the molecules that we can calculate analytically from the atomic coordinates. Image descriptors are a convenient tool here as they are not only visually intuitive for people but also suitable for many machine learning model designs. In all results, we focus on the atomic disks descriptor as it provides the most discrete predictions for the atomic positions and encodes the chemical information in the size of each disk, proportional to the covalent radius of the atom. The relative height of each atom is represented as brightness of the disk (background is zero), and atoms deeper than $1.2$ Å relative to the top atom are not considered.


    \subsection*{Machine learning}
    
    In this work, the structure discovery task is formulated as an image-to-image problem, for which we have developed and trained a machine learning model which translates the STM image into a descriptor by using an Attention U-Net-type model which utilizes an encoder-decoder architecture and an attention-gating mechanism \cite{Oktay2018}. On a conceptual level, the encoder compresses the information of the input image (STM image) into a latent space vector, which in turn is translated into a spatially larger representation (image descriptor) by the decoder. 
    The encoder consists of 4 convolutional blocks containing two 2D convolutional layers with a LeakyReLU$_{0.1}$ \cite{Xu2015} activation function in between, followed by batch normalization \cite{Ioffe2015} and a second LeakyReLU activation. Each of the convolutional blocks is preceded by a downscaling block which downsamples the image by using strided convolutions to increase the receptive field. The same activation and batch normalization policies are used in the downscaling. The decoder section consists of four blocks that include an upscaling block, an attention gate, and a convolutional block identical with the encoder. Upsampling is achieved by using a transposed 2D convolutional layer, and in the block, the same activation and batch normalization policies are used as in the encoder. The attention gate generates highlighted regions in the input for the model to focus on by using a type of skip connection where additional query signal comes from the upscaled feature map. The attention map is concatenated with the upsampled feature map and used as input for the convolutional block. The output layer of the model is a $1\times1$ convolutional layer followed by a ReLU activation (see full details in Fig. \ref{fig:attunet}).

    The ML model is implemented in PyTorch \cite{Paszke2019} and it is trained in a supervised setting for which the images were divided into training (180,000 images), validation (20,000) and testing (35,554) sets. The model was trained for 50 epochs and a batch size of 30 was used. The parameters of the model were optimized using the Adam optimizer \cite{Kingma2015} to minimize the mean squared error. During training, the simulated STM images are normalized to 0 mean and unit variance and we also add white noise and cut-outs to the images, representing electronic noise and tip artifacts to make the model not rely on pristine simulated images but instead force it to make predictions on imperfect images. After training, inference on the model can be done using true experimental images to obtain a prediction of the structure with very little computational cost.

    \section*{Results}

    \subsection*{Predictions on simulated images}
    
\begin{figure}
        \centering
        \includegraphics[width=1\linewidth]{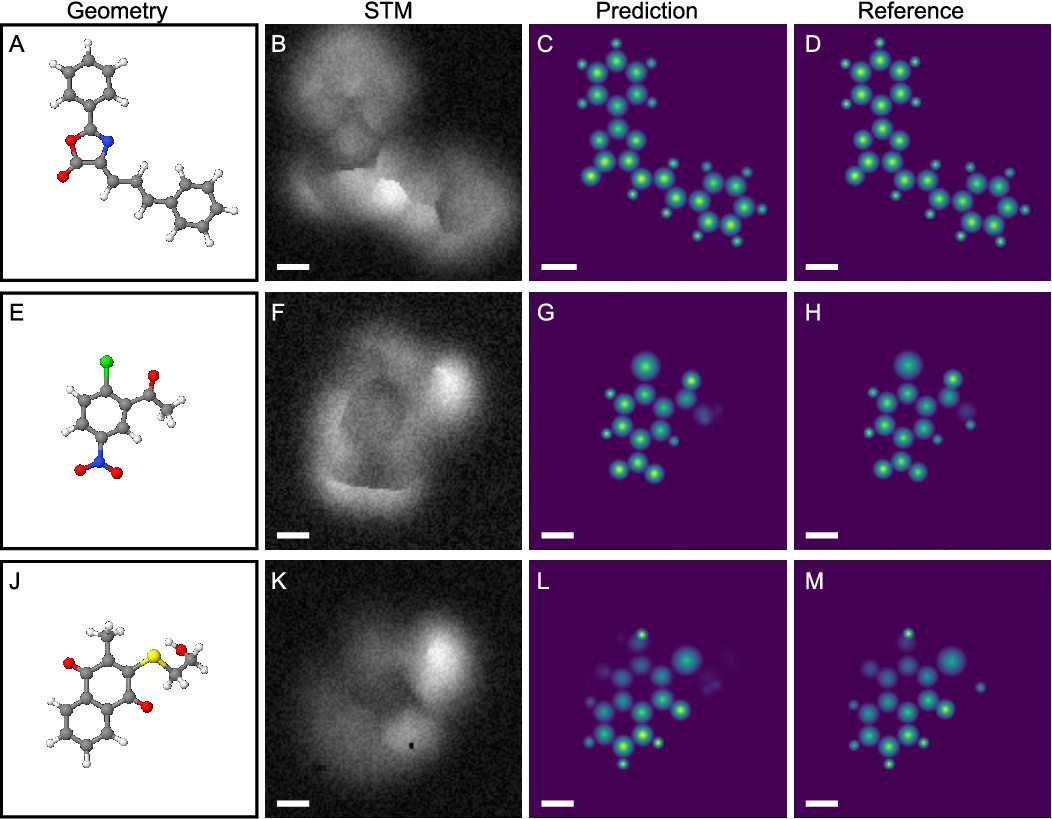}
        \caption{Example predictions of simulated STM images of (A-D) C$_{18}$H$_{13}$NO$_2$, (E-H) C$_8$H$_6$FNO$_3$ and (J-M) C$_{13}$H$_{12}$O$_3$S molecules from the testing set. Each molecule has a column for the geometry, STM image, predicted structure and reference. Scale bar is $2$ Å.}
        \label{fig:simulated}
    \end{figure}

    To benchmark our method and assess the improvement of prediction accuracy, we use the trained model to predict atomic structures from simulated images not included in the training data. The obvious benefit from analysing simulated predictions is that the atomic coordinates of the reference, and therefore the image descriptor, are known exactly. Fig.~\ref{fig:simulated} shows three example predictions from simulated images. The three examples have been taken at varying scanning heights and the sample molecules are different in size and structure, and they contain different chemical species and functional groups to demonstrate the versatility of the method. 
    
    Molecule 1 (Fig.~\ref{fig:simulated}A-D) is the largest of the three and it has two benzene groups and a 5-ring lactone group with an additional nitrogen heteroatom. The molecule is planar and all atoms are visible to the tip resulting in the carbon backbone skeleton being revealed with the exception of the benzene rings which are not prominent in the STM image (Fig.~\ref{fig:simulated}B). Even though the skeleton is not completely visible, the prediction (Fig.~\ref{fig:simulated}C) is very close to the reference (Fig.~\ref{fig:simulated}D). All atoms, including hydrogens which are not visible to the eye, are correctly located and the associated chemical identification distinguishes hydrogen atoms from carbon, nitrogen and oxygen atoms. It is also promising that the model correctly identifies the oxygen bonded to the 5-ring as not hydrogen and that it does not blindly add hydrogen atoms to nitrogen and oxygen atoms of the 5-ring. Overall, the mean absolute error relative to the range of values in the reference is $1.1$ \%. 
    
    The second example shows a smaller molecule containing a nitro group and a chlorine atom (Fig.~\ref{fig:simulated}E-H). The molecule is slightly tilted and it includes a non-planar part in which the branch containing a carbonyl group is bending downwards. This is projected in the reference descriptor by a barely visible carbon and two of the bonded hydrogen atoms are too deep to be considered (Fig.~\ref{fig:simulated}H). Again, the prediction resembles the reference very closely (Fig.~\ref{fig:simulated}G). The chlorine atom is distinguished from hydrogen atoms and the relative height of the oxygen in the carbonyl group is correctly identified. The location of the rightmost carbon atom is predicted correctly in the scanning plane but the disk is too bright, meaning that the vertical position prediction is wrong. Since the oxygen atom is approximately $1.0$ Å higher than the deepest carbon atom, the disk representing it is very bright in comparison and it appears to shadow the nearby deep carbon atom meaning it is not surprising that the model struggles most in this region. The relative error is $0.4$ \%.

    The final simulated example contains a sulphur atom, a hydroxy group and two carbon 6-rings, one of which has two carbonyl groups (Fig.~\ref{fig:simulated}J-M). In the STM image, the sulphur atom appears as the brightest area (Fig. \ref{fig:simulated}K). This molecule is particularly tricky for chemical identification due to the carbonyl groups which could theoretically be hydrogen atoms of a benzene ring, and the sulphur atom bridging the two carbon structures could be \textit{e.g.}, an oxygen atom forming an ether group. Regardless of the apparent challenges, the model predicts all these properties correctly (Fig.~\ref{fig:simulated}L). The disks representing oxygen atoms have a larger radius than the hydrogen atoms of the lower ring, and the sulphur atom has an even bigger radius distinguishing it from oxygen and carbon atoms. The carbon chain originating from the sulphur is not considered in the descriptor due to the depth threshold. Finally, all hydrogen atoms included in the reference are predicted and the tilting of the molecule is captured in the prediction. The relative error for this example is $0.6$ \%. Overall, the prediction accuracy is very good on simulated STM images and the trained model predicts atomic structures of different size, distinguishes various chemical species and even appears to predict seemingly hidden atoms. However, the final validation of the model has to be done on experimental data and to this end, we first consider four benchmark molecules.
    
    \subsection*{Experimental validation}

\begin{figure}
        \centering
        \includegraphics[width=1\linewidth]{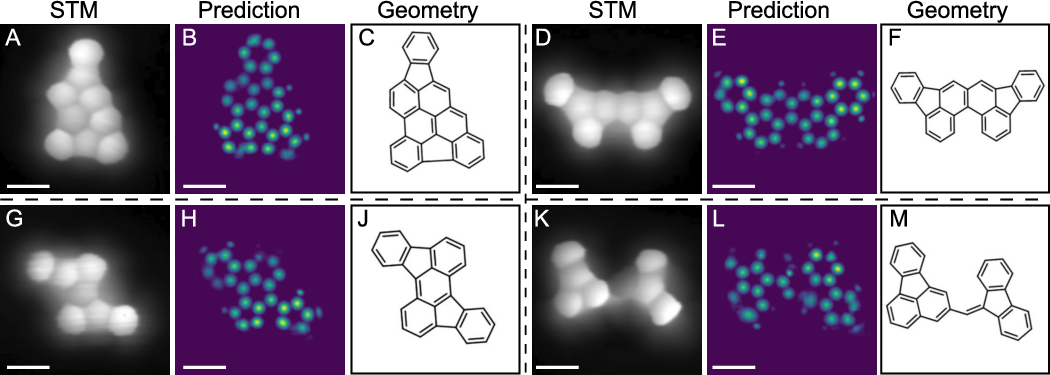}
                \caption{Predictions from four experimental STM images. (A, D, G, K) The first column of each example is the STM image used as input for ASD-STM. (B, E, H, L) The predictions are shown in the second column and (C, F, J, M) the third column contains the structural formula. Scale bar is $4$ Å. STM images and structural formulas adapted with permission from \cite{Song2020}. \copyright 2020 ACS.}
        \label{fig:exp1}
    \end{figure}

    The first experimental images we apply ASD-STM on are STM images obtained by Song \textit{et al.} \cite{Song2020} that reveal the molecular skeletons of four hydrocarbons in high detail making them excellent for benchmarking purposes (Fig.~\ref{fig:exp1}A, \ref{fig:exp1}D, \ref{fig:exp1}G, \ref{fig:exp1}K). The molecules consist of exclusively carbon and hydrogen atoms constructing five and six-membered rings barring the fourth example in which a carbon atom links two ring structures (Fig. \ref{fig:exp1}M). In all STM images, the rings appear quite distorted, and in many cases it is not immediately clear how many carbon atoms constitute the particular ring and therefore along with general structure discovery, we consider our objective to be to distinguish these rings. The first example consists of nine carbon-rings of which seven are six-membered. Most of the rings, seven out of nine, are correctly identified in the prediction (Fig. \ref{fig:exp1}B), characterization of one ring is unclear, and one ring is misclassified. The model also seems to have learned how rings of different size connect with each other. Additionally, the prediction suggests that the bottom and top parts of the molecule are slightly closer to the tip than the central parts but these predictions are difficult to validate.

    The second example is an STM image of a symmetric molecule with eight rings (Fig.~\ref{fig:exp1}D-F). The five-rings are fairly large in the image, and the rings in all ends of the molecule appear brighter than the center. The prediction is very good in terms of the carbon structure with all atoms predicted correctly (Fig.~\ref{fig:exp1}E). In this case, even many hydrogen atoms are included in the prediction. The third example we consider is smaller than the previous molecules (Fig.~\ref{fig:exp1}G-J) and again, the model performs well in predicting the atomic positions of carbon atoms. In this case, the bottom row of rings proposes a challenge for the model as the six-rings in the corners are misclassified as five-rings. A possible explanation for the wrong prediction is that the rings are almost circular and lack the characteristic corners. This is supported by the first example which exhibited similar features in the STM image, and where similar problems were encountered. It is also worth noting that the STM image contains considerable noise in the form of horizontal lines but the model succeeds in the prediction regardless. We attribute this to a successful augmentation process during training phase where noise and cut-outs were applied to the initially pristine simulated images.

    The final example is different from the others in that it includes two discrete ring structures that are connected by a carbon atom (Fig.~\ref{fig:exp1}K-M). It is clear that the model struggles with the molecule and does not predict the correct structure in this case. The rightmost part is mostly correct, although the model does not seem to fully recognize the rings, as some of the predicted atoms are blurry and slightly deformed. The leftmost part of the molecule has more errors, and only the central five-ring is correctly predicted (Fig.~\ref{fig:exp1}L). A general problem with many machine learning models is that their explainability is poor and it is often difficult to explain the reasons behind a certain prediction \cite{Linardatos2021}. One possible reason would be the particular structure of the molecule -- molecules containing two separate but connected carbon ring structures are rare in the data set, meaning the model has not had proper experience and thus fails in the prediction. Overall, the predictions correspond well to the molecular structures and in total 106 out of 116 ($91.4$ \%) carbon atoms and 23 out of 31 ($74.2$ \%) carbon rings are predicted correctly. 

\begin{figure}
        \centering
        \includegraphics[width=\linewidth]{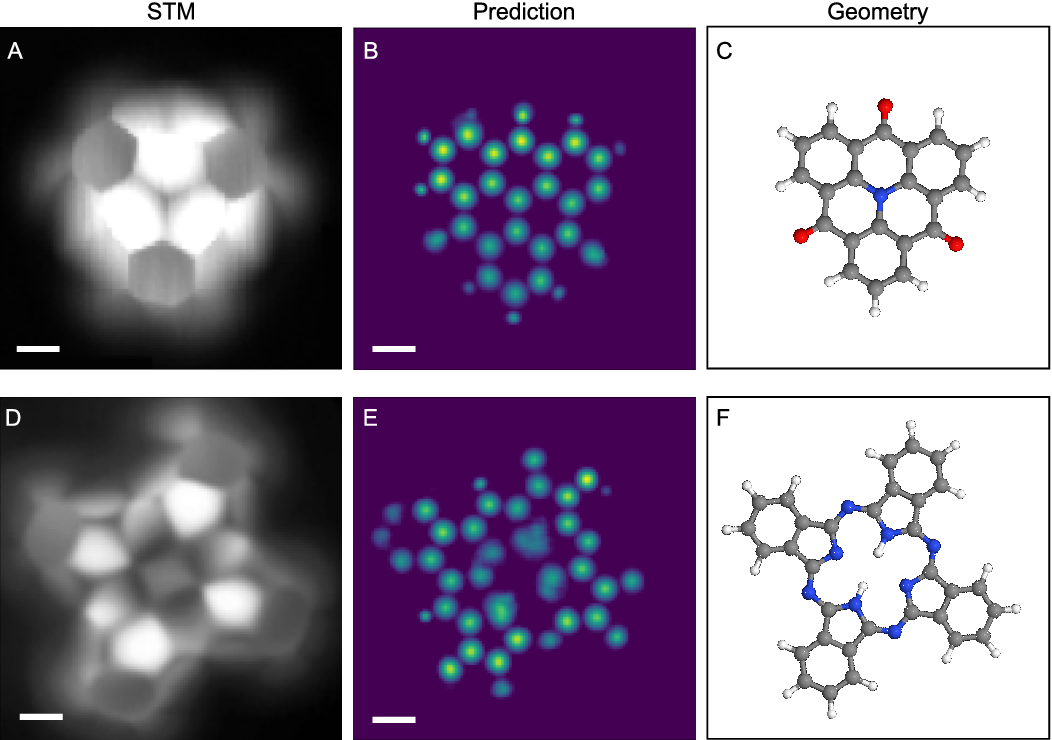}
        \caption{Prediction on experimental of image of the (A-C) TOAT (1,5,9-trioxo-13-azatriangulene) and (D-F) phthalocyanine (2H-Pc) molecules. For both molecules the columns contain, from left to right, the STM image, structure prediction and geometry of the molecule. Location of the pyrrolic hydrogen atoms in (F) as proposed in \cite{Neel2023}. Scale bar is $2$ Å. STM images in (A) reprinted with permission from \cite{Heijden2016} \copyright 2016 ACS and in (D) from \cite{Neel2023}  \copyright 2023 ACS.}
        \label{fig:exp2}
    \end{figure}

    Predictions on the hydrocarbon examples are promising but the molecular structures were rather simple, and we want to test the limits of the model on a more challenging example. To this end, we use STM images of the TOAT molecule (1,5,9-trioxo-13-azatriangulene) (Fig.~\ref{fig:exp2}A) \cite{Heijden2016} and phthalocyanine (2H-Pc) (Fig.~\ref{fig:exp2}D) \cite{Neel2023}. Predicting the atomic structure of TOAT is particularly interesting since the simulation software we used to generate the training data is known for not being able to properly reproduce experimental STM images of TOAT \cite{Krejci2017-ppstm}. Regardless, the prediction is in very good agreement with the true molecular structure (\ref{fig:exp2}B). The triangulene backbone is correctly identified and all oxygen atoms are distinguished from hydrogen atoms, indicated by larger disks in the prediction. In this example, even most hydrogen atoms are located with the only exception of the top right ring missing one hydrogen. The disk representing the center nitrogen seems to have the same radius as the neighboring carbon atoms suggesting an incorrect prediction but this classification is inconclusive. Definitive chemical identification has been explored previously for AFM \cite{Carracedo-Cosme2021, Oinonen2022a} but for STM it remains a future challenge.

    The most challenging example we test here is an STM image of 2H-Pc (Fig.~\ref{fig:exp2}D) and this prediction (Fig.~\ref{fig:exp2}E) reveals the limitations of our approach. 2H-Pc is a cyclic molecule consisting of four isoindole units connected by nitrogen atoms (Fig.~\ref{fig:exp2}F). In this example, a perfect prediction would reveal the configuration of hydrogen atoms in the central moiety. In the STM image, the inner five-rings are bright and clearly exhibit the pentagon-like motif whereas the outer six-rings are less pronounced. The nitrogen atoms connecting the units appear as cones decaying towards the center. In general, the molecular skeleton is not very prominent in the image and the signal originating from the electronic structure is intersecting significantly with the mechanical bending of the tip apex. It is apparent that these factors severely affect the ability of ASD-STM to make accurate predictions of the structure. In the prediction, the five-rings apart from the top left ring are correctly identified and all nitrogen atoms connecting the rings are located. The biggest errors are the missing halves of the six-rings and also the central hydrogens that are not included in the prediction. It is worth noting that a similar phenomenon was observed in the case of the fourth hydrocarbon (Fig. \ref{fig:exp1}L) where the top half of a ring was missing in the prediction.

    \section*{Discussion}
    
    While predictions on most experimental examples were good, it is clear there is room for improvement. Since the quality of data included in the training process is vital to the accuracy of predictions, rare structures affecting the accuracy should be discussed as it raises the important question of how to select the molecules in the data set. In the scope of SPM structure discovery, there are two main approaches to data set generation. The first option, which this work highlights, is to create a large, diverse and descriptive molecular data set of various chemical species and structures and to train a versatile model with the aim of predicting the structure of almost any small organic molecule. This approach has also been chosen in most previous sample characterization efforts (\textit{e.g.} \cite{Alldritt2020}, \cite{Carracedo-Cosme2022}) but recently another method has been used in ice structure discovery \cite{Tang2022, Priante2023} where instead of a diverse data set, a tailored data set is utilized and perfected to make very accurate predictions possible in a constrained problem domain. That is, if the goal was to only predict the geometries of different hydrocarbons or triangulene-based molecules, the model would benefit from a tailored data set with a heavy emphasis on such structures. As methods for quick generation of molecular structures to create vast molecular data sets are becoming more readily available \cite{Bilodeau2022, hoogeboom2022equivariant, daigavane2023symphony}, the balance and cost effectiveness of tailored versus diverse data sets should be the focus of future research.

    Second, ASD-STM has been trained on synthetic STM images of molecules in isolation, without the substrate and as such, the generated image data set should not be inferred as a source for accurate electronic structures of adsorbed organic molecules. Moreover, as we know that PPSTM cannot accurately reproduce STM images of the TOAT molecule, it was expected that ASD-STM would struggle with the experimental STM image of TOAT. However, the accuracy in the prediction was excellent which suggests that even though images in the training set are not entirely representative of true images in terms of electronic structure, they capture the characteristic sharp lines coming from CO tip bending accurately, and this seems to be critical for structure discovery. Also, synthetic AFM images of isolated molecules in general correspond well to experimental AFM images where the substrate is naturally present, and since for some samples it is possible to simultaneously gather STM and AFM signals, the possibility of incorporating AFM data into the training process should be explored.

    Third, we note that the appearance of the molecule in STM is sensitive to frontier orbitals of the CO tip and consequently to the tip height as at closer tip-sample distances the tip exhibits a strong $p$-wave character which diminishes to a more $s$-wave character at longer distances \cite{Neel2023}. The $s$-wave character is also increased at higher absolute bias voltages \cite{Pavlicek2013SymmetryTunneling}. In this work ASD-STM has been trained using a constant ratio of $s$ and $p$-wave contributions but the different orbital ratios should be accounted for if ASD-STM is used for a larger range of tip-sample distances or bias voltages. Finally, although it is inevitable that the accuracy of predictions on experimental images is worse than on simulated images, we will investigate this further in the future with particular focus in improving the robustness of the model against experimental noise and artifacts.    

    \section*{Conclusions}

    This work presents ASD-STM, a method for predicting the atomic structure of a sample molecule directly from a bond-resolved STM image. We present the workflow and the software required to generate a STM image data set and to train the ML model. Images included in the training set were synthesized by considering the molecules in isolation, and even though the electronic structure can be affected by the substrate, we showed that the approximation of molecules in isolation is reasonable for structure discovery on bond-resolved STM. The model was validated by applying it to experimental images of six different systems and on most samples the accuracy was good in terms of atomic structure -- additionally we achieved a qualitative level of chemical identification by distinguishing between hydrogen, carbon, and oxygen atoms. On the other hand, the example of 2H-Pc clearly demonstrated some of the challenges. The main limitation of the method is the reliance on the sharp submolecular features in the images which restrict the method to high quality images and to a range of short tip-sample distances. While the simulated images are augmented with noise and cut-outs during training, we anticipate that these augmentations are not the only difference between simulated and experimental images and advanced augmentation strategies should be explored to make the model more robust against experimental conditions. Also, we discussed possible further improvements to ASD-STM by varying the orbital contributions in the tunnelling calculation, by tailoring the composition of the initial data set of molecules to the problem at hand, and by including simultaneously gathered AFM data in the training process.
    
    Despite the challenges, ASD-STM is readily applicable to STM images of various small organic molecules and with this work, we open doors for more sophisticated methods to be developed for STM structure discovery. Finally, as bond-resolved STM images exhibit similar sub-molecular features as non-contact AFM images at a significantly reduced acquisition time, ASD-STM demonstrates a promising start for accelerating molecular structure discovery in SPM in general.

    \begin{acknowledgement}
    The authors acknowledge funding from the Academy of Finland (project no. 346824). A.S.F. was supported by the World Premier International Research Center Initiative (WPI), MEXT, Japan. L.K. acknowledges funding from the Finnish Cultural Foundation. The authors acknowledge the computational resources provided by the Aalto Science-IT project and CSC, Helsinki.
    \end{acknowledgement}

    \begin{suppinfo}
    Details of the machine learning model.
    \end{suppinfo}
    
    \clearpage
    \bibliography{refs}

\providecommand{\latin}[1]{#1}
\makeatletter
\providecommand{\doi}
  {\begingroup\let\do\@makeother\dospecials
  \catcode`\{=1 \catcode`\}=2 \doi@aux}
\providecommand{\doi@aux}[1]{\endgroup\texttt{#1}}
\makeatother
\providecommand*\mcitethebibliography{\thebibliography}
\csname @ifundefined\endcsname{endmcitethebibliography}  {\let\endmcitethebibliography\endthebibliography}{}
\begin{mcitethebibliography}{58}
\providecommand*\natexlab[1]{#1}
\providecommand*\mciteSetBstSublistMode[1]{}
\providecommand*\mciteSetBstMaxWidthForm[2]{}
\providecommand*\mciteBstWouldAddEndPuncttrue
  {\def\EndOfBibitem{\unskip.}}
\providecommand*\mciteBstWouldAddEndPunctfalse
  {\let\EndOfBibitem\relax}
\providecommand*\mciteSetBstMidEndSepPunct[3]{}
\providecommand*\mciteSetBstSublistLabelBeginEnd[3]{}
\providecommand*\EndOfBibitem{}
\mciteSetBstSublistMode{f}
\mciteSetBstMaxWidthForm{subitem}{(\alph{mcitesubitemcount})}
\mciteSetBstSublistLabelBeginEnd
  {\mcitemaxwidthsubitemform\space}
  {\relax}
  {\relax}

\bibitem[Binnig \latin{et~al.}(1982)Binnig, Rohrer, Gerber, and Weibel]{Binnig1982}
Binnig,~G.; Rohrer,~H.; Gerber,~C.; Weibel,~E. Tunneling through a controllable vacuum gap. \emph{Appl. Phys. Lett.} \textbf{1982}, \emph{40}, 178--180\relax
\mciteBstWouldAddEndPuncttrue
\mciteSetBstMidEndSepPunct{\mcitedefaultmidpunct}
{\mcitedefaultendpunct}{\mcitedefaultseppunct}\relax
\EndOfBibitem
\bibitem[Binnig \latin{et~al.}(1986)Binnig, Quate, and Gerber]{Binnig1986}
Binnig,~G.; Quate,~C.~F.; Gerber,~C. Atomic force microscope. \emph{Phys. Rev. Lett.} \textbf{1986}, \emph{56}, 930--933\relax
\mciteBstWouldAddEndPuncttrue
\mciteSetBstMidEndSepPunct{\mcitedefaultmidpunct}
{\mcitedefaultendpunct}{\mcitedefaultseppunct}\relax
\EndOfBibitem
\bibitem[Forsburg \latin{et~al.}(1990)Forsburg, Guarente, Johnson, Mcknight, Mitchell, Tjian, Buratowski, Hahn, Guarente, Sharp, Roeder, Sawadogo, Buratowski, Hahn, Sharp, Eisenmann, Dollard, and Winston]{Forsburg1990}
Forsburg,~S.; Guarente,~L.~A.; Johnson,~P. R.~.; Mcknight,~S.~A.; Mitchell,~P.~J.; Tjian,~R.; Buratowski,~S.~.; Hahn,~S.; Guarente,~L.; Sharp,~P. . M.~W.; Roeder,~R.~G.; Sawadogo,~M.; Buratowski,~S.; Hahn,~S.~.; Sharp,~P.; Eisenmann,~D.~M.; Dollard,~C.; Winston,~F. Atomic-scale imaging of DNA using scanning tunnelling microscopy. \emph{Nature} \textbf{1990}, \emph{346}, 294--296\relax
\mciteBstWouldAddEndPuncttrue
\mciteSetBstMidEndSepPunct{\mcitedefaultmidpunct}
{\mcitedefaultendpunct}{\mcitedefaultseppunct}\relax
\EndOfBibitem
\bibitem[Pawlak \latin{et~al.}(2019)Pawlak, Vilhena, Hinaut, Meier, Glatzel, Baratoff, Gnecco, Pérez, and Meyer]{Pawlak2019}
Pawlak,~R.; Vilhena,~J.~G.; Hinaut,~A.; Meier,~T.; Glatzel,~T.; Baratoff,~A.; Gnecco,~E.; Pérez,~R.; Meyer,~E. Conformations and cryo-force spectroscopy of spray-deposited single-strand DNA on gold. \emph{Nat. Commun.} \textbf{2019}, \emph{10}, 1--7\relax
\mciteBstWouldAddEndPuncttrue
\mciteSetBstMidEndSepPunct{\mcitedefaultmidpunct}
{\mcitedefaultendpunct}{\mcitedefaultseppunct}\relax
\EndOfBibitem
\bibitem[Dufrêne \latin{et~al.}(2017)Dufrêne, Ando, Garcia, Alsteens, Martinez-Martin, Engel, Gerber, and Müller]{Dufrene2017ImagingBiology}
Dufrêne,~Y.~F.; Ando,~T.; Garcia,~R.; Alsteens,~D.; Martinez-Martin,~D.; Engel,~A.; Gerber,~C.; Müller,~D.~J. Imaging modes of atomic force microscopy for application in molecular and cell biology. \emph{Nat. Nanotechnol.} \textbf{2017}, \emph{12}, 295--307\relax
\mciteBstWouldAddEndPuncttrue
\mciteSetBstMidEndSepPunct{\mcitedefaultmidpunct}
{\mcitedefaultendpunct}{\mcitedefaultseppunct}\relax
\EndOfBibitem
\bibitem[M{\"{u}}ller and Dufr{\^{e}}ne(2008)M{\"{u}}ller, and Dufr{\^{e}}ne]{Muller2008}
M{\"{u}}ller,~D.~J.; Dufr{\^{e}}ne,~Y.~F. {Atomic force microscopy as a multifunctional molecular toolbox in nanobiotechnology}. \emph{Nat. Nanotechnol.} \textbf{2008}, \emph{3}, 261--269\relax
\mciteBstWouldAddEndPuncttrue
\mciteSetBstMidEndSepPunct{\mcitedefaultmidpunct}
{\mcitedefaultendpunct}{\mcitedefaultseppunct}\relax
\EndOfBibitem
\bibitem[Gobbi \latin{et~al.}(2018)Gobbi, Orgiu, Samorì, Gobbi, Orgiu, and Samorì]{Gobbi2018}
Gobbi,~M.; Orgiu,~E.; Samorì,~P.; Gobbi,~M.; Orgiu,~.~E.; Samorì,~.~P. When 2D Materials Meet Molecules: Opportunities and Challenges of Hybrid Organic/Inorganic van der Waals Heterostructures. \emph{Adv. Mater.} \textbf{2018}, \emph{30}, 1706103\relax
\mciteBstWouldAddEndPuncttrue
\mciteSetBstMidEndSepPunct{\mcitedefaultmidpunct}
{\mcitedefaultendpunct}{\mcitedefaultseppunct}\relax
\EndOfBibitem
\bibitem[Sun \latin{et~al.}(2022)Sun, Silveira, Ma, Hasegawa, Matsumoto, Kera, Krejčí, Foster, and Kawai]{Sun2022}
Sun,~K.; Silveira,~O.~J.; Ma,~Y.; Hasegawa,~Y.; Matsumoto,~M.; Kera,~S.; Krejčí,~O.; Foster,~A.~S.; Kawai,~S. On-surface synthesis of disilabenzene-bridged covalent organic frameworks. \emph{Nat. Chem.} \textbf{2022}, \emph{15}, 136--142\relax
\mciteBstWouldAddEndPuncttrue
\mciteSetBstMidEndSepPunct{\mcitedefaultmidpunct}
{\mcitedefaultendpunct}{\mcitedefaultseppunct}\relax
\EndOfBibitem
\bibitem[Gross \latin{et~al.}(2018)Gross, Schuler, Pavliček, Fatayer, Majzik, Moll, Peña, and Meyer]{Gross2018}
Gross,~L.; Schuler,~B.; Pavliček,~N.; Fatayer,~S.; Majzik,~Z.; Moll,~N.; Peña,~D.; Meyer,~G. Atomic Force Microscopy for Molecular Structure Elucidation. \emph{Angew. Chem. Int. Ed.} \textbf{2018}, \emph{57}, 3888--3908\relax
\mciteBstWouldAddEndPuncttrue
\mciteSetBstMidEndSepPunct{\mcitedefaultmidpunct}
{\mcitedefaultendpunct}{\mcitedefaultseppunct}\relax
\EndOfBibitem
\bibitem[Gross \latin{et~al.}(2009)Gross, Mohn, Moll, Liljeroth, and Meyer]{Gross2009}
Gross,~L.; Mohn,~F.; Moll,~N.; Liljeroth,~P.; Meyer,~G. The chemical structure of a molecule resolved by atomic force microscopy. \emph{Science} \textbf{2009}, \emph{325}, 1110--1114\relax
\mciteBstWouldAddEndPuncttrue
\mciteSetBstMidEndSepPunct{\mcitedefaultmidpunct}
{\mcitedefaultendpunct}{\mcitedefaultseppunct}\relax
\EndOfBibitem
\bibitem[Bartels \latin{et~al.}(1997)Bartels, Meyer, and Rieder]{Bartels1997}
Bartels,~L.; Meyer,~G.; Rieder,~K.~H. Controlled vertical manipulation of single CO molecules with the scanning tunneling microscope: A route to chemical contrast. \emph{Appl. Phys. Lett.} \textbf{1997}, \emph{71}, 213--215\relax
\mciteBstWouldAddEndPuncttrue
\mciteSetBstMidEndSepPunct{\mcitedefaultmidpunct}
{\mcitedefaultendpunct}{\mcitedefaultseppunct}\relax
\EndOfBibitem
\bibitem[Zhong \latin{et~al.}(2020)Zhong, Li, Zhang, and Chi]{Zhong2020}
Zhong,~Q.; Li,~X.; Zhang,~H.; Chi,~L. Noncontact atomic force microscopy: Bond imaging and beyond. \emph{Surf. Sci. Rep.} \textbf{2020}, \emph{75}, 100509\relax
\mciteBstWouldAddEndPuncttrue
\mciteSetBstMidEndSepPunct{\mcitedefaultmidpunct}
{\mcitedefaultendpunct}{\mcitedefaultseppunct}\relax
\EndOfBibitem
\bibitem[Jelinek(2017)]{Jelinek2017}
Jelinek,~P. High resolution SPM imaging of organic molecules with functionalized tips. \emph{J. Condens. Matter Phys.} \textbf{2017}, \emph{29}, 343002\relax
\mciteBstWouldAddEndPuncttrue
\mciteSetBstMidEndSepPunct{\mcitedefaultmidpunct}
{\mcitedefaultendpunct}{\mcitedefaultseppunct}\relax
\EndOfBibitem
\bibitem[Pavliček \latin{et~al.}(2017)Pavliček, Mistry, Majzik, Moll, Meyer, Fox, and Gross]{Pavlicek2017}
Pavliček,~N.; Mistry,~A.; Majzik,~Z.; Moll,~N.; Meyer,~G.; Fox,~D.~J.; Gross,~L. Synthesis and characterization of triangulene. \emph{Nat. Nanotechnol.} \textbf{2017}, \emph{12}, 308--311\relax
\mciteBstWouldAddEndPuncttrue
\mciteSetBstMidEndSepPunct{\mcitedefaultmidpunct}
{\mcitedefaultendpunct}{\mcitedefaultseppunct}\relax
\EndOfBibitem
\bibitem[Seibel \latin{et~al.}(2023)Seibel, Fittolani, Mirhosseini, Wu, Rauschenbach, Anggara, Seeberger, Delbianco, Kühne, Schlickum, and Kern]{Seibel2023}
Seibel,~J.; Fittolani,~G.; Mirhosseini,~H.; Wu,~X.; Rauschenbach,~S.; Anggara,~K.; Seeberger,~P.~H.; Delbianco,~M.; Kühne,~T.~D.; Schlickum,~U.; Kern,~K. Visualizing Chiral Interactions in Carbohydrates Adsorbed on Au(111) by High-Resolution STM Imaging. \emph{Angew. Chem. Int. Ed.} \textbf{2023}, \emph{62}, e202305733\relax
\mciteBstWouldAddEndPuncttrue
\mciteSetBstMidEndSepPunct{\mcitedefaultmidpunct}
{\mcitedefaultendpunct}{\mcitedefaultseppunct}\relax
\EndOfBibitem
\bibitem[Vilas-Varela \latin{et~al.}(2023)Vilas-Varela, Romero-Lara, Vegliante, Calupitan, Martínez, Meyer, Uriarte-Amiano, Friedrich, Wang, Schulz, Koval, Sandoval-Salinas, Casanova, Corso, Artacho, Peña, Pascual, Vilas-Varela, Martínez, Peña, Romero-Lara, Vegliante, Meyer, Uriarte-Amiano, Friedrich, Wang, Schulz, Koval, Artacho, Pascual, Sandoval-Salinas, and Casanova]{Vilas-Varela2023}
Vilas-Varela,~M.; Romero-Lara,~F.; Vegliante,~A.; Calupitan,~J.~P.; Martínez,~A.; Meyer,~L.; Uriarte-Amiano,~U.; Friedrich,~N.; Wang,~D.; Schulz,~F.; Koval,~N.~E.; Sandoval-Salinas,~M.~E.; Casanova,~D.; Corso,~M.; Artacho,~E.; Peña,~D.; Pascual,~J.~I.; Vilas-Varela,~M.; Martínez,~A.; Peña,~D. \latin{et~al.}  On-Surface Synthesis and Characterization of a High-Spin Aza-[5]-Triangulene. \emph{Angew. Chem. Int. Ed.} \textbf{2023}, \emph{62}, e202307884\relax
\mciteBstWouldAddEndPuncttrue
\mciteSetBstMidEndSepPunct{\mcitedefaultmidpunct}
{\mcitedefaultendpunct}{\mcitedefaultseppunct}\relax
\EndOfBibitem
\bibitem[Martinez-Castro \latin{et~al.}(2022)Martinez-Castro, Bolat, Fan, Werner, Arefi, Esat, Sundermeyer, Wagner, Gottfried, Temirov, Ternes, and Tautz]{Martinez-Castro2022}
Martinez-Castro,~J.; Bolat,~R.; Fan,~Q.; Werner,~S.; Arefi,~H.~H.; Esat,~T.; Sundermeyer,~J.; Wagner,~C.; Gottfried,~J.~M.; Temirov,~R.; Ternes,~M.; Tautz,~F.~S. Disentangling the electronic structure of an adsorbed graphene nanoring by scanning tunneling microscopy. \emph{Commun. Mater.} \textbf{2022}, \emph{3}, 1--9\relax
\mciteBstWouldAddEndPuncttrue
\mciteSetBstMidEndSepPunct{\mcitedefaultmidpunct}
{\mcitedefaultendpunct}{\mcitedefaultseppunct}\relax
\EndOfBibitem
\bibitem[Song \latin{et~al.}(2020)Song, Guo, Li, Li, Haketa, Telychko, Su, Lyu, Qiu, Fang, Peng, Li, Wu, Li, Su, Koh, Wu, Maeda, Zhang, and Lu]{Song2020}
Song,~S.; Guo,~N.; Li,~X.; Li,~G.; Haketa,~Y.; Telychko,~M.; Su,~J.; Lyu,~P.; Qiu,~Z.; Fang,~H.; Peng,~X.; Li,~J.; Wu,~X.; Li,~Y.; Su,~C.; Koh,~M.~J.; Wu,~J.; Maeda,~H.; Zhang,~C.; Lu,~J. Real-Space Imaging of a Single-Molecule Monoradical Reaction. \emph{J. Am. Chem. Soc.} \textbf{2020}, \emph{142}, 13550--13557\relax
\mciteBstWouldAddEndPuncttrue
\mciteSetBstMidEndSepPunct{\mcitedefaultmidpunct}
{\mcitedefaultendpunct}{\mcitedefaultseppunct}\relax
\EndOfBibitem
\bibitem[Weiss \latin{et~al.}(2010)Weiss, Wagner, Temirov, and Tautz]{Weiss2010}
Weiss,~C.; Wagner,~C.; Temirov,~R.; Tautz,~F.~S. Direct imaging of intermolecular bonds in scanning tunneling microscopy. \emph{J. Am. Chem. Soc.} \textbf{2010}, \emph{132}, 11864--11865\relax
\mciteBstWouldAddEndPuncttrue
\mciteSetBstMidEndSepPunct{\mcitedefaultmidpunct}
{\mcitedefaultendpunct}{\mcitedefaultseppunct}\relax
\EndOfBibitem
\bibitem[Gross \latin{et~al.}(2011)Gross, Moll, Mohn, Curioni, Meyer, Hanke, and Persson]{Gross2011ptip}
Gross,~L.; Moll,~N.; Mohn,~F.; Curioni,~A.; Meyer,~G.; Hanke,~F.; Persson,~M. High-Resolution Molecular Orbital Imaging Using a p-Wave STM Tip. \emph{Phys. Rev. Lett.} \textbf{2011}, \emph{107}, 86101\relax
\mciteBstWouldAddEndPuncttrue
\mciteSetBstMidEndSepPunct{\mcitedefaultmidpunct}
{\mcitedefaultendpunct}{\mcitedefaultseppunct}\relax
\EndOfBibitem
\bibitem[Néel and Kröger(2023)Néel, and Kröger]{Neel2023}
Néel,~N.; Kröger,~J. Orbital and Skeletal Structure of a Single Molecule on a Metal Surface Unveiled by Scanning Tunneling Microscopy. \emph{J. Phys. Chem. Lett.} \textbf{2023}, \emph{14}, 3946--3952\relax
\mciteBstWouldAddEndPuncttrue
\mciteSetBstMidEndSepPunct{\mcitedefaultmidpunct}
{\mcitedefaultendpunct}{\mcitedefaultseppunct}\relax
\EndOfBibitem
\bibitem[Alldritt \latin{et~al.}(2020)Alldritt, Hapala, Oinonen, Urtev, Krejci, Canova, Kannala, Schulz, Liljeroth, and Foster]{Alldritt2020}
Alldritt,~B.; Hapala,~P.; Oinonen,~N.; Urtev,~F.; Krejci,~O.; Canova,~F.~F.; Kannala,~J.; Schulz,~F.; Liljeroth,~P.; Foster,~A.~S. Automated structure discovery in atomic force microscopy. \emph{Sci. Adv.} \textbf{2020}, \emph{6}, eaay6913\relax
\mciteBstWouldAddEndPuncttrue
\mciteSetBstMidEndSepPunct{\mcitedefaultmidpunct}
{\mcitedefaultendpunct}{\mcitedefaultseppunct}\relax
\EndOfBibitem
\bibitem[Carracedo-Cosme \latin{et~al.}(2021)Carracedo-Cosme, Romero-Muñiz, and Pérez]{Carracedo-Cosme2021}
Carracedo-Cosme,~J.; Romero-Muñiz,~C.; Pérez,~R. A Deep Learning Approach for Molecular Classification Based on AFM Images. \emph{Nanomater.} \textbf{2021}, \emph{11}, 1658\relax
\mciteBstWouldAddEndPuncttrue
\mciteSetBstMidEndSepPunct{\mcitedefaultmidpunct}
{\mcitedefaultendpunct}{\mcitedefaultseppunct}\relax
\EndOfBibitem
\bibitem[Oinonen \latin{et~al.}(2022)Oinonen, Kurki, Ilin, and Foster]{Oinonen2022a}
Oinonen,~N.; Kurki,~L.; Ilin,~A.; Foster,~A.~S. Molecule graph reconstruction from atomic force microscope images with machine learning. \emph{MRS Bull.} \textbf{2022}, \emph{47}, 895--905\relax
\mciteBstWouldAddEndPuncttrue
\mciteSetBstMidEndSepPunct{\mcitedefaultmidpunct}
{\mcitedefaultendpunct}{\mcitedefaultseppunct}\relax
\EndOfBibitem
\bibitem[Oinonen \latin{et~al.}(2022)Oinonen, Xu, Alldritt, Canova, Urtev, Cai, Krejčí, Kannala, Liljeroth, and Foster]{Oinonen2022b}
Oinonen,~N.; Xu,~C.; Alldritt,~B.; Canova,~F.~F.; Urtev,~F.; Cai,~S.; Krejčí,~O.; Kannala,~J.; Liljeroth,~P.; Foster,~A.~S. Electrostatic Discovery Atomic Force Microscopy. \emph{ACS Nano} \textbf{2022}, \emph{16}, 89--97\relax
\mciteBstWouldAddEndPuncttrue
\mciteSetBstMidEndSepPunct{\mcitedefaultmidpunct}
{\mcitedefaultendpunct}{\mcitedefaultseppunct}\relax
\EndOfBibitem
\bibitem[Tang \latin{et~al.}(2022)Tang, Song, Qin, Tian, Wu, Jiang, Cao, and Xu]{Tang2022}
Tang,~B.; Song,~Y.; Qin,~M.; Tian,~Y.; Wu,~Z.~W.; Jiang,~Y.; Cao,~D.; Xu,~L. Machine learning aided atomic structure identification of interfacial ionic hydrates from AFM images. \emph{Natl. Sci. Rev.} \textbf{2022}, \emph{10}, nwac282\relax
\mciteBstWouldAddEndPuncttrue
\mciteSetBstMidEndSepPunct{\mcitedefaultmidpunct}
{\mcitedefaultendpunct}{\mcitedefaultseppunct}\relax
\EndOfBibitem
\bibitem[Carracedo-Cosme \latin{et~al.}(2023)Carracedo-Cosme, Romero-Muñiz, Pou, and Pérez]{Carracedo-Cosme2023}
Carracedo-Cosme,~J.; Romero-Muñiz,~C.; Pou,~P.; Pérez,~R. Molecular Identification from AFM Images Using the IUPAC Nomenclature and Attribute Multimodal Recurrent Neural Networks. \emph{ACS Appl. Mater. Interfaces} \textbf{2023}, \emph{15}, 22692--22704\relax
\mciteBstWouldAddEndPuncttrue
\mciteSetBstMidEndSepPunct{\mcitedefaultmidpunct}
{\mcitedefaultendpunct}{\mcitedefaultseppunct}\relax
\EndOfBibitem
\bibitem[Wang \latin{et~al.}(2020)Wang, Li, Hao, Li, Zou, Cai, Wang, You, and Zhai]{Wang2020}
Wang,~C.; Li,~H.; Hao,~Z.; Li,~X.; Zou,~C.; Cai,~P.; Wang,~Y.; You,~Y.~Z.; Zhai,~H. Machine learning identification of impurities in the STM images. \emph{Chin. Physics B} \textbf{2020}, \emph{29}, 116805\relax
\mciteBstWouldAddEndPuncttrue
\mciteSetBstMidEndSepPunct{\mcitedefaultmidpunct}
{\mcitedefaultendpunct}{\mcitedefaultseppunct}\relax
\EndOfBibitem
\bibitem[Yuan \latin{et~al.}(2023)Yuan, Zhu, Lu, Zheng, Jiang, and Sun]{Yuan2023}
Yuan,~S.; Zhu,~Z.; Lu,~J.; Zheng,~F.; Jiang,~H.; Sun,~Q. Applying a Deep-Learning-Based Keypoint Detection in Analyzing Surface Nanostructures. \emph{Molecules} \textbf{2023}, \emph{28}, 5387\relax
\mciteBstWouldAddEndPuncttrue
\mciteSetBstMidEndSepPunct{\mcitedefaultmidpunct}
{\mcitedefaultendpunct}{\mcitedefaultseppunct}\relax
\EndOfBibitem
\bibitem[Choudhary \latin{et~al.}(2021)Choudhary, Garrity, Camp, Kalinin, Vasudevan, Ziatdinov, and Tavazza]{Choudhary2021}
Choudhary,~K.; Garrity,~K.~F.; Camp,~C.; Kalinin,~S.~V.; Vasudevan,~R.; Ziatdinov,~M.; Tavazza,~F. Computational scanning tunneling microscope image database. \emph{Sci Data} \textbf{2021}, \emph{8}, 1--9\relax
\mciteBstWouldAddEndPuncttrue
\mciteSetBstMidEndSepPunct{\mcitedefaultmidpunct}
{\mcitedefaultendpunct}{\mcitedefaultseppunct}\relax
\EndOfBibitem
\bibitem[Chen \latin{et~al.}(2022)Chen, Aapro, Kipnis, Ilin, Liljeroth, and Foster]{Chen2022}
Chen,~I.~J.; Aapro,~M.; Kipnis,~A.; Ilin,~A.; Liljeroth,~P.; Foster,~A.~S. Precise atom manipulation through deep reinforcement learning. \emph{Nat. Commun.} \textbf{2022}, \emph{13}, 1--8\relax
\mciteBstWouldAddEndPuncttrue
\mciteSetBstMidEndSepPunct{\mcitedefaultmidpunct}
{\mcitedefaultendpunct}{\mcitedefaultseppunct}\relax
\EndOfBibitem
\bibitem[Rashidi and Wolkow(2018)Rashidi, and Wolkow]{Rashidi2018}
Rashidi,~M.; Wolkow,~R.~A. Autonomous Scanning Probe Microscopy in Situ Tip Conditioning through Machine Learning. \emph{ACS Nano} \textbf{2018}, \emph{12}, 5185--5189\relax
\mciteBstWouldAddEndPuncttrue
\mciteSetBstMidEndSepPunct{\mcitedefaultmidpunct}
{\mcitedefaultendpunct}{\mcitedefaultseppunct}\relax
\EndOfBibitem
\bibitem[Wang \latin{et~al.}(2021)Wang, Zhu, Blackwell, and Fischer]{Wang2021}
Wang,~S.; Zhu,~J.; Blackwell,~R.; Fischer,~F.~R. Automated tip conditioning for scanning tunneling spectroscopy. \emph{J. Phys. Chem. A} \textbf{2021}, \emph{125}, 1384--1390\relax
\mciteBstWouldAddEndPuncttrue
\mciteSetBstMidEndSepPunct{\mcitedefaultmidpunct}
{\mcitedefaultendpunct}{\mcitedefaultseppunct}\relax
\EndOfBibitem
\bibitem[Alldritt \latin{et~al.}(2022)Alldritt, Urtev, Oinonen, Aapro, Kannala, Liljeroth, and Foster]{Alldritt2022}
Alldritt,~B.; Urtev,~F.; Oinonen,~N.; Aapro,~M.; Kannala,~J.; Liljeroth,~P.; Foster,~A.~S. Automated tip functionalization via machine learning in scanning probe microscopy. \emph{Comp. Phys. Comp.} \textbf{2022}, \emph{273}, 108258\relax
\mciteBstWouldAddEndPuncttrue
\mciteSetBstMidEndSepPunct{\mcitedefaultmidpunct}
{\mcitedefaultendpunct}{\mcitedefaultseppunct}\relax
\EndOfBibitem
\bibitem[Gross(2011)]{Gross2011review}
Gross,~L. Recent advances in submolecular resolution with scanning probe microscopy. \emph{Nat. Chem.} \textbf{2011}, \emph{3}, 273--278\relax
\mciteBstWouldAddEndPuncttrue
\mciteSetBstMidEndSepPunct{\mcitedefaultmidpunct}
{\mcitedefaultendpunct}{\mcitedefaultseppunct}\relax
\EndOfBibitem
\bibitem[Blum \latin{et~al.}(2009)Blum, Gehrke, Hanke, Havu, Havu, Ren, Reuter, and Scheffler]{Blum2009}
Blum,~V.; Gehrke,~R.; Hanke,~F.; Havu,~P.; Havu,~V.; Ren,~X.; Reuter,~K.; Scheffler,~M. Ab initio molecular simulations with numeric atom-centered orbitals. \emph{Comp. Phys. Comm.} \textbf{2009}, \emph{180}, 2175--2196\relax
\mciteBstWouldAddEndPuncttrue
\mciteSetBstMidEndSepPunct{\mcitedefaultmidpunct}
{\mcitedefaultendpunct}{\mcitedefaultseppunct}\relax
\EndOfBibitem
\bibitem[Havu \latin{et~al.}(2009)Havu, Blum, Havu, and Scheffler]{Havu2009}
Havu,~V.; Blum,~V.; Havu,~P.; Scheffler,~M. Efficient O(N) integration for all-electron electronic structure calculation using numeric basis functions. \emph{J. Comput. Phys.} \textbf{2009}, \emph{228}, 8367--8379\relax
\mciteBstWouldAddEndPuncttrue
\mciteSetBstMidEndSepPunct{\mcitedefaultmidpunct}
{\mcitedefaultendpunct}{\mcitedefaultseppunct}\relax
\EndOfBibitem
\bibitem[Perdew \latin{et~al.}(1996)Perdew, Burke, and Ernzerhof]{Perdew1996}
Perdew,~J.~P.; Burke,~K.; Ernzerhof,~M. Generalized Gradient Approximation Made Simple. \emph{Phys. Rev. Lett.} \textbf{1996}, \emph{77}, 3865\relax
\mciteBstWouldAddEndPuncttrue
\mciteSetBstMidEndSepPunct{\mcitedefaultmidpunct}
{\mcitedefaultendpunct}{\mcitedefaultseppunct}\relax
\EndOfBibitem
\bibitem[Krejčí \latin{et~al.}(2017)Krejčí, Hapala, Ondráček, and Jelínek]{Krejci2017-ppstm}
Krejčí,~O.; Hapala,~P.; Ondráček,~M.; Jelínek,~P. Principles and simulations of high-resolution STM imaging with a flexible tip apex. \emph{Phys. Rev. B} \textbf{2017}, \emph{95}\relax
\mciteBstWouldAddEndPuncttrue
\mciteSetBstMidEndSepPunct{\mcitedefaultmidpunct}
{\mcitedefaultendpunct}{\mcitedefaultseppunct}\relax
\EndOfBibitem
\bibitem[Bardeen(1961)]{Bardeen1961}
Bardeen,~J. Tunnelling from a Many-Particle Point of View. \emph{Phys. Rev. Lett.} \textbf{1961}, \emph{6}, 57\relax
\mciteBstWouldAddEndPuncttrue
\mciteSetBstMidEndSepPunct{\mcitedefaultmidpunct}
{\mcitedefaultendpunct}{\mcitedefaultseppunct}\relax
\EndOfBibitem
\bibitem[Torre \latin{et~al.}(2017)Torre, Švec, Foti, Krejčí, Hapala, Garcia-Lekue, Frederiksen, Zbořil, Arnau, Vázquez, and Jelínek]{DeLaTorre2017}
Torre,~B. D.~L.; Švec,~M.; Foti,~G.; Krejčí,~O.; Hapala,~P.; Garcia-Lekue,~A.; Frederiksen,~T.; Zbořil,~R.; Arnau,~A.; Vázquez,~H.; Jelínek,~P. Submolecular Resolution by Variation of the Inelastic Electron Tunneling Spectroscopy Amplitude and its Relation to the AFM/STM Signal. \emph{Phys. Rev. Lett.} \textbf{2017}, \emph{119}, 166001\relax
\mciteBstWouldAddEndPuncttrue
\mciteSetBstMidEndSepPunct{\mcitedefaultmidpunct}
{\mcitedefaultendpunct}{\mcitedefaultseppunct}\relax
\EndOfBibitem
\bibitem[Hapala \latin{et~al.}(2014)Hapala, Kichin, Wagner, Tautz, Temirov, and Jelínek]{Hapala2014-ppafm1}
Hapala,~P.; Kichin,~G.; Wagner,~C.; Tautz,~F.~S.; Temirov,~R.; Jelínek,~P. Mechanism of high-resolution STM/AFM imaging with functionalized tips. \emph{Phys. Rev. B} \textbf{2014}, \emph{90}, 1--9\relax
\mciteBstWouldAddEndPuncttrue
\mciteSetBstMidEndSepPunct{\mcitedefaultmidpunct}
{\mcitedefaultendpunct}{\mcitedefaultseppunct}\relax
\EndOfBibitem
\bibitem[Hapala \latin{et~al.}(2014)Hapala, Temirov, Tautz, and Jelínek]{Hapala2014-ppafm2}
Hapala,~P.; Temirov,~R.; Tautz,~F.~S.; Jelínek,~P. Origin of high-resolution IETS-STM images of organic molecules with functionalized tips. \emph{Phys. Rev. Lett.} \textbf{2014}, \emph{113}, 226101\relax
\mciteBstWouldAddEndPuncttrue
\mciteSetBstMidEndSepPunct{\mcitedefaultmidpunct}
{\mcitedefaultendpunct}{\mcitedefaultseppunct}\relax
\EndOfBibitem
\bibitem[Giessibl(2001)]{Giessibl2001}
Giessibl,~F.~J. A direct method to calculate tip-sample forces from frequency shifts in frequency-modulation atomic force microscopy. \emph{Appl. Phys. Lett.} \textbf{2001}, \emph{78}, 123--125\relax
\mciteBstWouldAddEndPuncttrue
\mciteSetBstMidEndSepPunct{\mcitedefaultmidpunct}
{\mcitedefaultendpunct}{\mcitedefaultseppunct}\relax
\EndOfBibitem
\bibitem[Oktay \latin{et~al.}(2018)Oktay, Schlemper, Folgoc, Lee, Heinrich, Misawa, Mori, McDonagh, Hammerla, Kainz, Glocker, and Rueckert]{Oktay2018}
Oktay,~O.; Schlemper,~J.; Folgoc,~L.~L.; Lee,~M.; Heinrich,~M.; Misawa,~K.; Mori,~K.; McDonagh,~S.; Hammerla,~N.~Y.; Kainz,~B.; Glocker,~B.; Rueckert,~D. Attention U-Net: Learning Where to Look for the Pancreas. 2018; arXiv:1804.03999\relax
\mciteBstWouldAddEndPuncttrue
\mciteSetBstMidEndSepPunct{\mcitedefaultmidpunct}
{\mcitedefaultendpunct}{\mcitedefaultseppunct}\relax
\EndOfBibitem
\bibitem[Xu \latin{et~al.}(2015)Xu, Wang, Kong, Chen, and Li]{Xu2015}
Xu,~B.; Wang,~N.; Kong,~H.; Chen,~T.; Li,~M. Empirical Evaluation of Rectified Activations in Convolutional Network. 2015; arXiv:1505.00853v2\relax
\mciteBstWouldAddEndPuncttrue
\mciteSetBstMidEndSepPunct{\mcitedefaultmidpunct}
{\mcitedefaultendpunct}{\mcitedefaultseppunct}\relax
\EndOfBibitem
\bibitem[Ioffe and Szegedy(2015)Ioffe, and Szegedy]{Ioffe2015}
Ioffe,~S.; Szegedy,~C. Batch Normalization: Accelerating Deep Network Training by Reducing Internal Covariate Shift. \emph{ICML} \textbf{2015}, \emph{1}, 448--456\relax
\mciteBstWouldAddEndPuncttrue
\mciteSetBstMidEndSepPunct{\mcitedefaultmidpunct}
{\mcitedefaultendpunct}{\mcitedefaultseppunct}\relax
\EndOfBibitem
\bibitem[Paszke \latin{et~al.}(2019)Paszke, Gross, Massa, Lerer, Bradbury, Chanan, Killeen, Lin, Gimelshein, Antiga, Desmaison, Köpf, Yang, DeVito, Raison, Tejani, Chilamkurthy, Steiner, Fang, Bai, and Chintala]{Paszke2019}
Paszke,~A.; Gross,~S.; Massa,~F.; Lerer,~A.; Bradbury,~J.; Chanan,~G.; Killeen,~T.; Lin,~Z.; Gimelshein,~N.; Antiga,~L.; Desmaison,~A.; Köpf,~A.; Yang,~E.; DeVito,~Z.; Raison,~M.; Tejani,~A.; Chilamkurthy,~S.; Steiner,~B.; Fang,~L.; Bai,~J. \latin{et~al.}  PyTorch: An Imperative Style, High-Performance Deep Learning Library. \emph{NeurIPS} \textbf{2019}, \emph{32}\relax
\mciteBstWouldAddEndPuncttrue
\mciteSetBstMidEndSepPunct{\mcitedefaultmidpunct}
{\mcitedefaultendpunct}{\mcitedefaultseppunct}\relax
\EndOfBibitem
\bibitem[Kingma and Ba(2015)Kingma, and Ba]{Kingma2015}
Kingma,~D.~P.; Ba,~J.~L. Adam: A method for stochastic optimization. 2015; arXiv:1412.6980v9\relax
\mciteBstWouldAddEndPuncttrue
\mciteSetBstMidEndSepPunct{\mcitedefaultmidpunct}
{\mcitedefaultendpunct}{\mcitedefaultseppunct}\relax
\EndOfBibitem
\bibitem[Linardatos \latin{et~al.}(2021)Linardatos, Papastefanopoulos, and Kotsiantis]{Linardatos2021}
Linardatos,~P.; Papastefanopoulos,~V.; Kotsiantis,~S. Explainable AI: A Review of Machine Learning Interpretability Methods. \emph{Entropy} \textbf{2021}, \emph{23}, 1--45\relax
\mciteBstWouldAddEndPuncttrue
\mciteSetBstMidEndSepPunct{\mcitedefaultmidpunct}
{\mcitedefaultendpunct}{\mcitedefaultseppunct}\relax
\EndOfBibitem
\bibitem[Heijden \latin{et~al.}(2016)Heijden, Hapala, Rombouts, Lit, Smith, Mutombo, Švec, Jelinek, and Swart]{Heijden2016}
Heijden,~N. J. V.~D.; Hapala,~P.; Rombouts,~J.~A.; Lit,~J. V.~D.; Smith,~D.; Mutombo,~P.; Švec,~M.; Jelinek,~P.; Swart,~I. Characteristic Contrast in $\Delta$fmin Maps of Organic Molecules Using Atomic Force Microscopy. \emph{ACS Nano} \textbf{2016}, \emph{10}, 8517--8525\relax
\mciteBstWouldAddEndPuncttrue
\mciteSetBstMidEndSepPunct{\mcitedefaultmidpunct}
{\mcitedefaultendpunct}{\mcitedefaultseppunct}\relax
\EndOfBibitem
\bibitem[Carracedo-Cosme \latin{et~al.}(2022)Carracedo-Cosme, Iz, Pou, and Pérez]{Carracedo-Cosme2022}
Carracedo-Cosme,~J.; Iz,~C. R.-M.; Pou,~P.; Pérez,~R. QUAM-AFM: A Free Database for Molecular Identification by Atomic Force Microscopy. \emph{J. Chem. Inf. Model} \textbf{2022}, \emph{62}, 38\relax
\mciteBstWouldAddEndPuncttrue
\mciteSetBstMidEndSepPunct{\mcitedefaultmidpunct}
{\mcitedefaultendpunct}{\mcitedefaultseppunct}\relax
\EndOfBibitem
\bibitem[Priante \latin{et~al.}(2023)Priante, Oinonen, Tian, Guan, Xu, Cai, Liljeroth, Jiang, and Foster]{Priante2023}
Priante,~F.; Oinonen,~N.; Tian,~Y.; Guan,~D.; Xu,~C.; Cai,~S.; Liljeroth,~P.; Jiang,~Y.; Foster,~A.~S. Structure discovery in Atomic Force Microscopy imaging of ice. 2023; arXiv:2310.17161v1\relax
\mciteBstWouldAddEndPuncttrue
\mciteSetBstMidEndSepPunct{\mcitedefaultmidpunct}
{\mcitedefaultendpunct}{\mcitedefaultseppunct}\relax
\EndOfBibitem
\bibitem[Bilodeau \latin{et~al.}(2022)Bilodeau, Jin, Jaakkola, Barzilay, and Jensen]{Bilodeau2022}
Bilodeau,~C.; Jin,~W.; Jaakkola,~T.; Barzilay,~R.; Jensen,~K.~F. Generative models for molecular discovery: Recent advances and challenges. \emph{Wiley Interdiscip. Rev. Comput. Mol. Sci.} \textbf{2022}, \emph{12}, e1608\relax
\mciteBstWouldAddEndPuncttrue
\mciteSetBstMidEndSepPunct{\mcitedefaultmidpunct}
{\mcitedefaultendpunct}{\mcitedefaultseppunct}\relax
\EndOfBibitem
\bibitem[Hoogeboom \latin{et~al.}(2022)Hoogeboom, Satorras, Vignac, and Welling]{hoogeboom2022equivariant}
Hoogeboom,~E.; Satorras,~V.~G.; Vignac,~C.; Welling,~M. Equivariant Diffusion for Molecule Generation in 3D. 2022; arXiv:2203.17003\relax
\mciteBstWouldAddEndPuncttrue
\mciteSetBstMidEndSepPunct{\mcitedefaultmidpunct}
{\mcitedefaultendpunct}{\mcitedefaultseppunct}\relax
\EndOfBibitem
\bibitem[Daigavane \latin{et~al.}(2023)Daigavane, Kim, Geiger, and Smidt]{daigavane2023symphony}
Daigavane,~A.; Kim,~S.; Geiger,~M.; Smidt,~T. Symphony: Symmetry-Equivariant Point-Centered Spherical Harmonics for Molecule Generation. 2023; arXiv:2311.16199\relax
\mciteBstWouldAddEndPuncttrue
\mciteSetBstMidEndSepPunct{\mcitedefaultmidpunct}
{\mcitedefaultendpunct}{\mcitedefaultseppunct}\relax
\EndOfBibitem
\bibitem[Pavli{\v{c}}ek \latin{et~al.}(2013)Pavli{\v{c}}ek, Swart, Niedenf{\"{u}}hr, Meyer, and Repp]{Pavlicek2013SymmetryTunneling}
Pavli{\v{c}}ek,~N.; Swart,~I.; Niedenf{\"{u}}hr,~J.; Meyer,~G.; Repp,~J. {Symmetry dependence of vibration-assisted tunneling}. \emph{Phys. Rev. Lett.} \textbf{2013}, \emph{110}, 136101\relax
\mciteBstWouldAddEndPuncttrue
\mciteSetBstMidEndSepPunct{\mcitedefaultmidpunct}
{\mcitedefaultendpunct}{\mcitedefaultseppunct}\relax
\EndOfBibitem
\end{mcitethebibliography}
    
\end{document}


\section*{Machine learning model}

\begin{figure}
        \centering
        \includegraphics[width=1.0\linewidth]{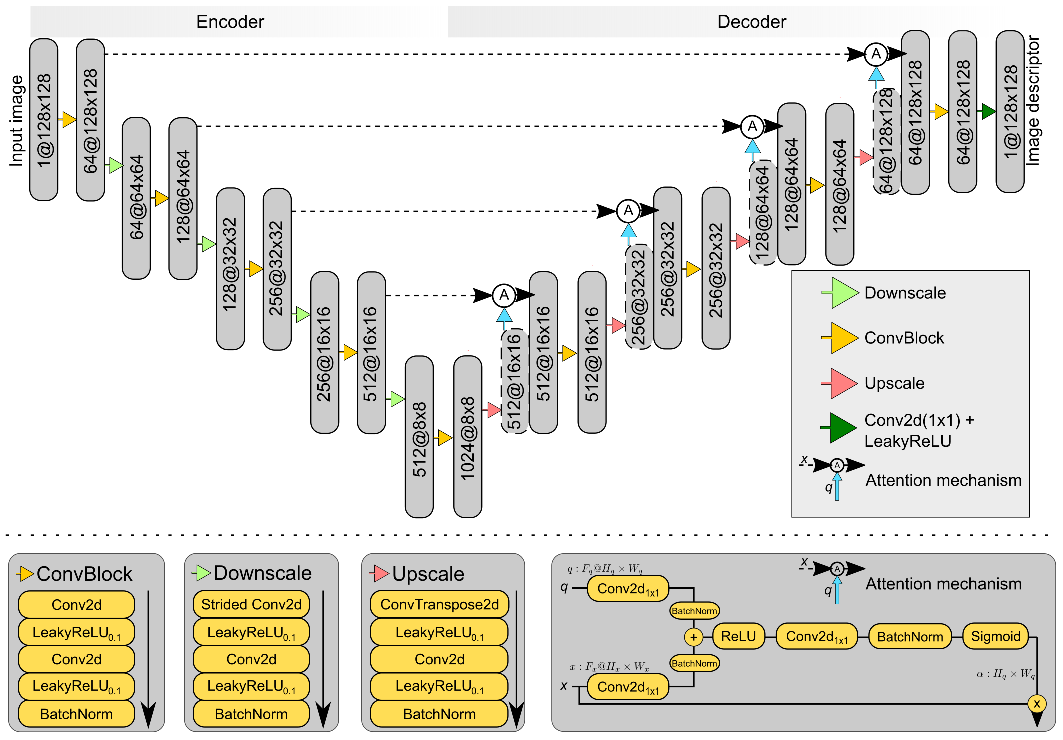}
        \caption{Diagram of the machine learning model used in ASD-STM. Different operations are denoted by colored arrows and all procedures are visualized in their respective boxes. The numbers in each block represent the number of channels and the spatial size of the feature map at every step of the model (F@W$\times$H). Input and output are $128\times128$ images with one channel.}
        \label{fig:attunet}
    \end{figure}

    ASD-STM uses an Attention U-Net-type model originally introduced in \cite{Oktay2018}. Overview of the model is shown in Fig. \ref{fig:attunet}. It uses the U-Net architecture with an attention mechanism instead of the standard skip connection. In the encoder, the model gradually shrinks the spatial size of the input while increasing the number of channels. The process is reversed in the decoder and in the end, the spatial size of the output is the same as in the input. The implementation of the Attention U-Net in ASD-STM uses a different number of layers and channels and also processes 2 dimensional input data whereas the original model uses 3 dimensional input. Here, the query signal is also obtained from the already upsampled feature map meaning that resampling is not needed in the attention gate as in the original model.

    The input layer of the model is a 2D convolutional block (Fig. \ref{fig:attunet}) with 32 channels. After the input layer, at every step of the encoder there is a downscaling block and a 2D convolutional block with increasing filter dimension and decreasing spatial size. In the decoder, the output of the ConvBlock is upscaled and used as the query in the attention mechanism, where the other input comes from the skip connection. The output of the attention mechanism is concatenated with the upscaled tensor and used as input for a ConvBlock. This process is repeated four times gradually decreasing the filter dimension and increasing the spatial size of the feature map, until the dimension in $64@128\times128$ ($F@W\times H$, assuming $128\times128$ input). Finally, there is a $1\times1$ convolutional layer and a ReLU activation (Eq. \ref{eq:relu}) which reduce the filter dimension to 1 and translate the output values to $x \geq 0$ to match the values in the descriptor.

    A ConvBlock consists of two $3\times3$ 2D convolutional layers, first of which increases (encoder) or decreases (decoder) the number of filters. In between, there is a LeakyReLU$_{0.1}$ (Eq. \ref{eq:leaky}) activation and finally a 2D batch normalization layer. A downscaling block is otherwise identical but the first convolutional layer is replaced by a $4\times4$ 2D convolutional layer with 2-stride to reduce the spatial size of the feature map. This operation is used instead of maximum pooling that was used in the original model. Similarly, in an upscaling block the first layer is replaced with a $4\times4$ 2D transposed convolutional layer with 2-stride to expand the feature map spatially. The final piece in the model is the attention mechanism which receives two inputs -- one from the skip connection $x$ and one from the query $q$. A $1\times1$ convolution and a batch normalization are applied to both to make the tensors one dimensional in the filter dimension, and the tensors are combined in a summing operation activated with a ReLU function to form the attention weighting vector $\alpha$. Subsequently, there is a second $1\times1$ convolution, a batch normalization and a sigmoid activation (Eq. \ref{eq:sigmoid}) to translate the vector into range $\alpha\in\left[0,1\right]$. Finally, the skip connection $x$ is multiplied with $\alpha$ to highlight salient regions in the feature map. The total number of parameters in the model is 51,588,013.

    The activation functions used in this model are defined as follows:
    \begin{align}
        \text{ReLU}(x) &= \max\left(0, x\right) \label{eq:relu} \\
        \text{LeakyReLU}(x)_\alpha &= \max\left(\alpha x, x\right),\quad \alpha \leq 1 \label{eq:leaky} \\
        \sigma(x) &= \frac{1}{1+e^{-x}} \label{eq:sigmoid}
    \end{align}

    \clearpage
    \bibliography{refs}